\documentstyle[twoside,fleqn,espcrc2,epsf]{article}

\def\half{{\textstyle{1\over2}}}
\def\){\right)} 
\def\({\left(} 
\def\]{\right]} 
\def\[{\left[}
\newcommand{\eqn}[1]{\label{eq:#1}}
\newcommand{\refeq}[1]{(\ref{eq:#1})}

\newcommand{\Eq}{Eq.~\refeq} 

\newcommand{\beq}{\begin{eqnarray}}
\newcommand{\eeq}{\end{eqnarray}}

\newcommand{\mcal}[1]{{\mathcal #1}}
\newcommand{\makefigreal}[4]{\begin{figure}[t] 
                           \centerline{\epsfysize=#3 in \epsfbox{#2}} 
			\vspace{-0.2truein}
                           \caption{#4 \label{#1}} 
			\vspace{-0.2truein}
                         \end{figure}}

\begin{document}

\title{
Partially Quenched QCD with Non-Degenerate Dynamical Quarks 
\thanks{Supported by DOE contract DE-FG03-96ER40956.}
}

\author{S. Sharpe 
\address{Physics Department, Box 351560,
University of Washington, Seattle, WA 98195-1560, USA} 
and N. Shoresh${}^{a}$\thanks{Speaker}}
      
\begin{abstract}
We discuss the importance of using partially quenched theories
with three degenerate quarks for extrapolating to QCD, and present 
some relevant results from chiral perturbation theory.

\end{abstract}

\maketitle

Simulations of lattice QCD have, to date, largely used
the quenched (Q), or partially quenched (PQ), approximations.
These approximations introduce unphysical artifacts, some of which can
be analyzed using chiral perturbation theory (ChPT)~\cite{BG,SS,BG2}.
In particular, it has been found that the quenched theory
has unphysical singularities in the chiral limit,
and that these persist in the PQ theory, albeit in weakened 
form~\cite{BG2,SS2,GL}.
These results serve as a warning against relying on ChPT when straying too far from 
unquenched theories in which valence and dynamical masses coincide.

In this talk we discuss a more constructive aspect of partially quenched
QCD. We focus on PQ theories with three light dynamical quarks,
where ``light'' means that the theory
can be studied using ChPT.
Such theories have the same quark complement as QCD, 
although with differing masses.
When simulating such theories it is possible to push to much 
lighter valence quarks than dynamical quarks, and thus to
map out a ``rectangular'' region in the space of theories as sketched
in Fig.~\ref{mspace}. Our aim is to make use of the extra information
contained in the PQ data to learn about unquenched theories,
and in particular about real QCD. The key point is
that, as long as one stays far enough away from the $m_{V}=0$ axis,
ChPT reliably predicts
the functional dependence on valence and dynamical quark masses,
and so provides a unified description of both the PQ and fully unquenched 
regions of the parameter space.
\makefigreal{mspace}{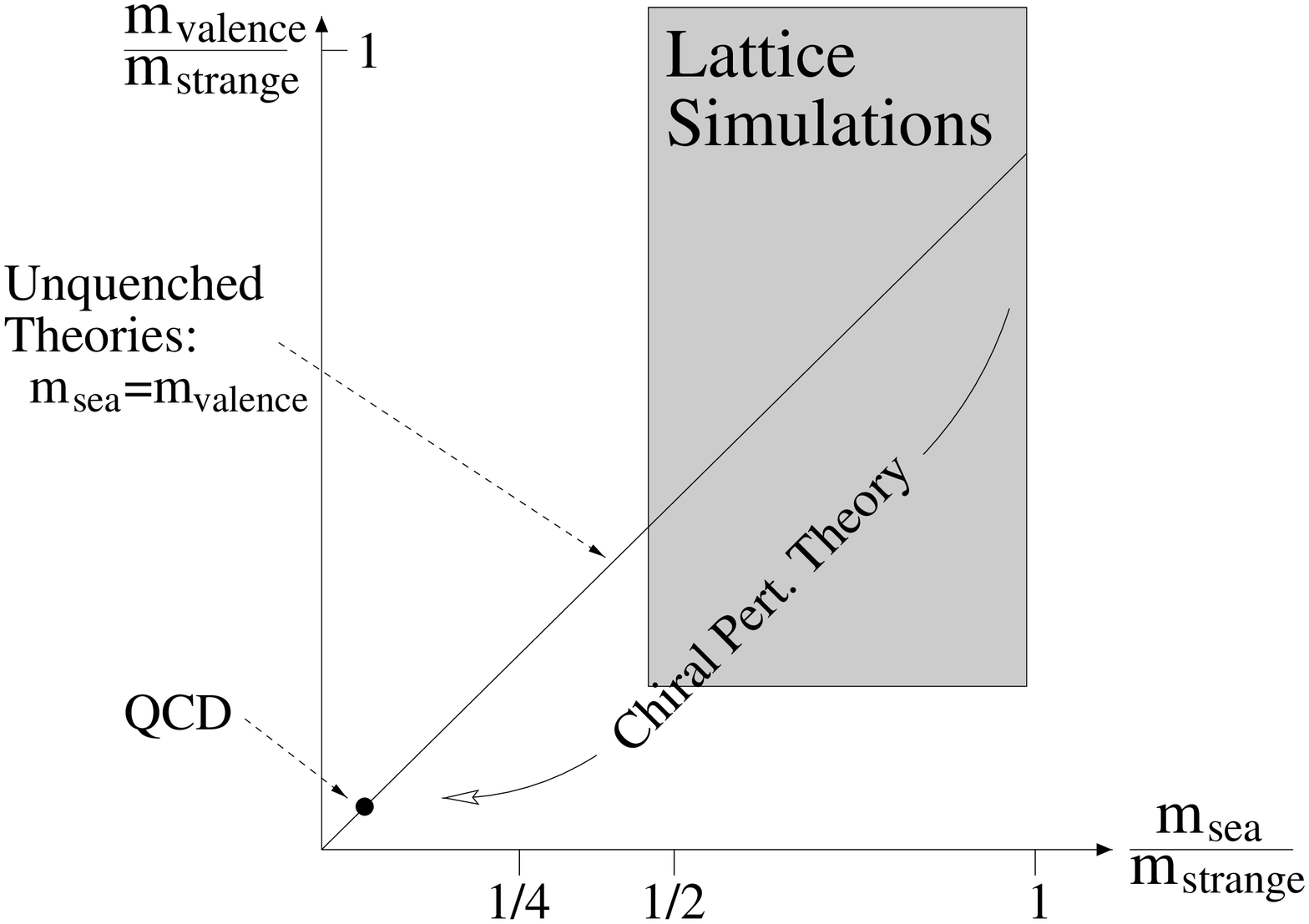}{2}{Schematic representation of the space of
PQ theories.}

An observation central to this work is that the unknown coefficients
appearing in the chiral Lagrangian for PQ theories depend only on
the number of light dynamical quarks, 
and not on their masses~\cite{SS2}\footnote{%
This result is also implicit in the work of Ref.~\cite{BG2}.}.
One way of viewing the extra input that PQ theories provide 
is that they allow a cleaner separation of the non-leading,
$O(p^4)$ coefficients\footnote{%
Which we refer to as Gasser-Leutwyler
(GL) coefficients.
}.
This is because some GL coefficients appear multiplied only by valence masses, 
others only by dynamical masses (see \Eq{eqM2}).

The picture that emerges is that, by
simulating in a region of moderately light valence and dynamical quarks, 
{\em and varying independently the masses of both}, one can determine
all the coefficients necessary to extrapolate to QCD.
This viewpoint has also been stressed in Ref.~\cite{CKN}
in the context of determining the up quark mass.
The resulting accuracy is limited by the order to which one calculates in
ChPT.

An important question is whether it is necessary in such a program of
calculations to consider non-degenerate dynamical quarks.
For the quantities we consider here---pion masses and decay 
constants---degenerate dynamical quarks suffice at one-loop in ChPT,
because the expressions contain only the sum of the dynamical 
masses\footnote{%
This appears to hold for all observables at one-loop order.}.
Simulations with
non-degenerate dynamical quarks are necessary for more accurate
chiral extrapolations, since mass differences enter in two-loop ChPT results.
Such simulations are also attractive because they bring one closer
to QCD, and thus reduce the dependence on ChPT.
They may also be advantageous from a computational point
of view---since a dynamical strange quark is relatively cheap,
one may approach
more closely to QCD by having two lighter dynamical quarks than three.

For these reasons we think it likely that simulations with
non-degenerate dynamical quarks will be done, and thus as a first
step in our program we have calculated the one-loop 
expressions for the masses and decay constants of the pions and kaons
in a theory with three dynamical quarks, only two of which are
degenerate. This generalizes Refs.~\cite{SS2,GL}, which calculated these
quantities with any number of degenerate dynamical quarks. 

\bigskip
The partially quenched approximation can be described in terms of a Lagrangian
by adding  a number of commuting spin-1/2 (ghost) fields~\cite{morel}. 
For each valence quark there is a corresponding ghost quark,
and when their masses are equal
the valence quark loops cancel against their ghost counterparts.
The dynamical quarks, which do not have corresponding ghosts,
give rise to the quark loops that are kept in the simulation.
The complete mass matrix is 
\beq
M=\mbox{diag}\(
m_{V1},m_{V2};m_{S1},m_{S2},m_{S3};\widetilde{m_{V1}},\widetilde{m_{V2}}\),
\nonumber
\eeq 
where $V$ refers to valence, $S$ to sea or dynamical, and a tilde
indicates ghost quantities. In our theory $m_{Vi}=\widetilde{m_{Vi}}$, 
and $m_{S1}=m_{S2}$. For baryons one would need to include an additional
valence quark and its corresponding ghost.

The approximate $\mathrm{SU}_L(3)\otimes\mathrm{SU}_R(3)$ 
chiral symmetry of QCD generalizes to the graded group
$\mathrm{SU}_L(5|2)\otimes\mathrm{SU}_R(5|2)$ in the PQ theory~\cite{BG2}.
Assuming that this larger symmetry is broken spontaneously down to
its vector subgroup, one finds a large multiplet of pseudo-Goldstone
particles which includes both bosons (quark-antiquark and ghost-antighost)
and fermions (quark-antighost). Collecting these into a matrix, $\Phi$,
and defining $\Sigma=\exp(2i\Phi/\mathrm{f})$ 
and $\chi=2\mu M$, the chiral Lagrangian
describing the low-momentum interactions of these particles is~\cite{BG2,SS2}
\beq
\lefteqn{\mcal{L}=(\mathrm{f}^2/4)\mathrm{str} 
               \( \partial\Sigma\partial\Sigma^\dagger \) 
           - (\mathrm{f}^2/4) \mathrm{str} 
               \( \chi\Sigma^\dagger+\Sigma\chi \) }
\nonumber \\
&& +\alpha_\Phi\half\(\partial\Phi_0\)^2+\half m_0^2\Phi_0^2 
\nonumber \\ 
&& +      L_4\,\mathrm{str}\(
\partial\Sigma\partial\Sigma^\dagger \) 
\mathrm{str}\( \chi\Sigma^\dagger+\Sigma\chi \) 
\nonumber \\
&&                 +L_5\,\mathrm{str}\( 
     \(\partial\Sigma\partial\Sigma^\dagger \) 
     \(\chi\Sigma^\dagger+\Sigma\chi \) 
                                     \)  
\nonumber \\
&&
-L_6\,\mathrm{str} 
                 \( (\chi\Sigma^\dagger+\Sigma\chi )^2 \) 
\nonumber \\
&&
-L_8\,\mathrm{str} 
                 \( \chi\Sigma^\dagger\chi\Sigma^\dagger+\Sigma\chi\Sigma\chi 
                   \) 
          +\ldots  \eqn{PQChL} 
\eeq 
This form is the same as that for QCD except for the replacement
of traces by supertraces, and for the appearance of the super-$\eta'$
field $\Phi_0\propto \mathrm{str}(\Phi)$.  
As in standard ChPT there are unknown constants at each order:
$\mathrm{f}$, $\mu$, $\alpha_\Phi$ and $m_0$ at leading order, 
and the GL coefficients $L_i$ at one-loop
(of which only those that are relevant are displayed).
Each term in \Eq{PQChL} can also be multiplied by an arbitrary even
function of $\Phi_0$. These additional functions do not, however, enter
into our final results and are not shown.

The QCD chiral Lagrangian is completely contained in \Eq{PQChL}, 
and can be obtained from it by setting valence and dynamical quark masses 
equal~\cite{BG2}.
Moreover, since all quark mass dependence appears explicitly in \Eq{PQChL}, 
and all the unknown coefficients are independent of masses, 
these coefficients are {\em the same} for the partially 
quenched theory and full QCD. 

Using \Eq{PQChL}, we calculate the masses and  decay constants 
of the Goldstone particles to one-loop in ChPT~\cite{ShSh}.
The results are functions of valence and dynamical masses: 
$M^2(m_{V1},m_{V2};m_{S1},m_{S3})$ and $F(m_{V1},m_{V2};m_{S1},m_{S3})$.
Unquenched pion-like properties are obtained for
$m_{V1}=m_{V2}=m_{S1}$,  whereas kaon-like properties are obtained for
$m_{V1}=m_{S2}$ and $m_{V2}=m_{S3}$. 
The results contain a simple term analytic in quark masses,
e.g.
\beq
\lefteqn{ M^2 = \mu (m_{V1} + m_{V2})
- (8 \mu^2/f^2)\left[ \right. }\nonumber \\
&& 2(L_4-2 L_6) (m_{V1}+m_{V2})(2m_{S1}+m_{S3})  \nonumber \\
&& \left. + (L_5-2L_8) (m_{V1}+m_{V2})^2 \right]\,,\label{eq:eqM2}
\eeq
and a long and unilluminating chiral logarithm arising from loops. 
Note that the dependence of $M^2$
on $(m_{V1}\!+\!m_{V2})^2$ gives the combination $L_5\!-\!2 L_8$,
the value of which is needed to determine the physical $m_{up}$~\cite{CKN}.

We use the full one-loop expressions to study issues that
arise in the extrapolation towards QCD. 
To do this we use
a simplification noted in Ref.~\cite{SS2}: for three dynamical quarks,
the super-$\eta'$ has a mass comparable to the physical $\eta'$
and can be integrated out. The resulting forms are
independent of the parameters $\alpha_\Phi$ and $m_0$ and the
extra functions of $\Phi_0$ implicit in \Eq{PQChL}.
They thus depend only on parameters that are present in the
QCD chiral Lagrangian.
We then find values for $\mathrm{f}$, and for
the GL coefficients,
such that there are choices for 
the ``physical'' up and strange quark  masses, denoted $m_{up}$ and $m_{st}$, 
which reproduce correctly the experimental values
of $f_\pi$, $f_K$, $m_\pi$ and $m_K$.  
There is a range of allowed GL coefficients---we select
representative samples by the additional criterion that chiral perturbation
theory should be converging.

Our first study concerns the importance of including non-analytic 
terms when extrapolating to QCD.
We produce ``fake'' data 
by calculating $M^2$ and $F$ (using the full one-loop forms) on
a preset grid of quark masses.  
The data is then fit to the one-loop 
analytic form predicted by ChPT (e.g.~\Eq{eqM2}), 
excluding the chiral logarithms.
This fit is then used to extrapolate to the physical values for the
ratios $m_\pi/f_K$  and  $m_K/f_K$. 
This leaves $f_\pi/f_K$ as a ``prediction'', which 
can  be compared  with the exact  value.
We find that, for grids with masses above $\sim m_{st}/8$, this
procedure leads to $10-15\%$ errors in $f_\pi/f_K$.
This is true also if one extrapolates solely using unquenched data.
We conclude that the extra curvature due to the chiral logarithms
is substantial,
and needs to be incorporated in the extrapolation.

Most current PQ simulations use two dynamical quarks, 
which can be viewed as the infinite strange quark mass 
limit of the three flavor theory.
An important issue is how large an error one makes by excluding the
dynamical strange quark. ChPT cannot be used to 
calculate this effect, since it cannot be applied if $m_{st}$ is too heavy. 
Nevertheless, by varying $m_{st}$ within the range of validity of ChPT, 
one can obtain an estimate of the error.
Fig.~\ref{deg} shows how the (normalized) ratio of $m_K/f_K$ varies as 
the mass of the third dynamical quark is increased from $m_{up}$ 
(i.e. three degenerate dynamical quarks) 
to $m_{st}$ (the physical point). Note that
the valence quark masses are unchanged.
We see an effect of $\sim 10\%$---large enough to suggest 
that inclusion of the dynamical strange quark is important.
     \makefigreal{deg}{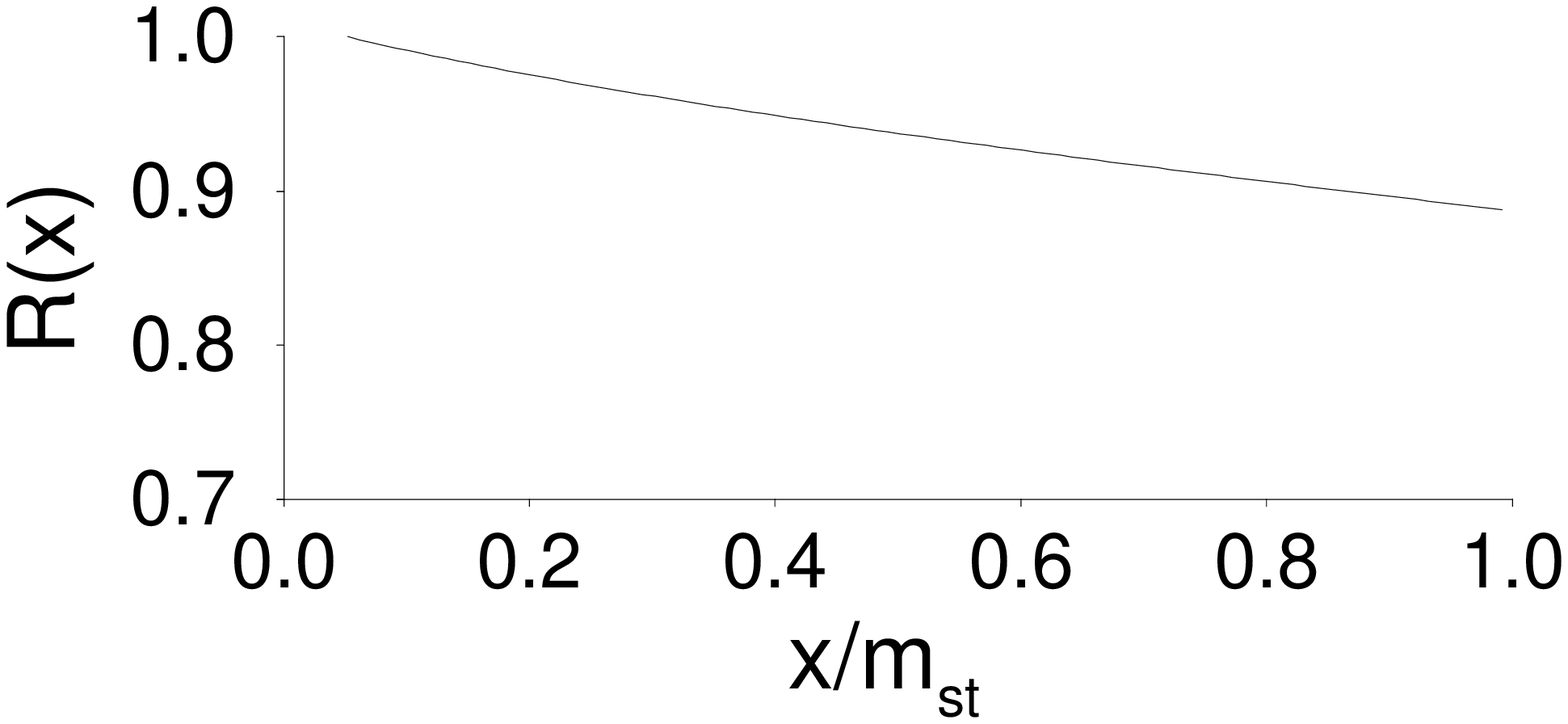}{1.4}{Effect of
     varying the dynamical strange quark
     mass. Quark masses are $m_{V1}=m_{up}$,
     $m_{V2}=m_{st}$, $m_{S1}=m_{S2}=m_{up}$ and
     $m_{S3}=x$.
     $R(x)=[M_K(x)/f_K(x)]/[M_K(m_{up})/f_K(m_{up})]$.}

More details concerning these exercises, as well as the one-loop
results for three non-degenerate quarks of arbitrary mass,
will be given in~\cite{ShSh}.


\begin{thebibliography}{9}

\bibitem{BG}
C.W.~Bernard and M.F.L.~Golterman, \\
Phys.~Rev.~D46, 853 (1992).
\bibitem{SS}
S.~Sharpe, Phys.~Rev.~D46, 3146 (1992).
\bibitem{BG2}
C.W.~Bernard and M.F.L.~Golterman, \\ Phys.~Rev.~D49, 486 (1994).
\bibitem{SS2}
S.~Sharpe, Phys.~Rev.~D56, 7052 (1997).
\bibitem{GL}
M.F.L.~Golterman and K.-C.~Leung, \\ Phys.~Rev.~D57, 5703 (1998).
\bibitem{CKN}
A.~Cohen, D.~Kaplan and A.~Nelson,\\ hep-lat/9909091.
\bibitem{morel}
A.~Morel, J.~Phys. (Paris) 48, 1111 (1987).
\bibitem{ShSh}
S.~Sharpe and N.~Shoresh, in preparation.

\end{thebibliography}
\end{document}